\documentclass[sort&compress]{elsarticle}

\usepackage{graphicx,amsmath}
\usepackage{enumerate}

\def\be{\begin{equation}}
\def\ee{\end{equation}}

\begin{document}

\begin{frontmatter}
\title{{\bf On parton number fluctuations at various stages of the
rapidity evolution}
}

\author[1]{A. H. Mueller}
\author[2]{S. Munier\corref{cor}}
\cortext[cor]{Corresponding author.}
\ead{Stephane.Munier@polytechnique.edu}

\address[1]{Department of Physics, Columbia University, New York, USA}
\address[2]{Centre de physique th\'eorique, \'Ecole Polytechnique, 
CNRS, Palaiseau, France}

\begin{abstract}
Starting with the interpretation of parton evolution with rapidity
as a branching-diffusion process,
we describe the different kinds of 
fluctuations 
of the density of partons
which affect the properties of 
QCD scattering amplitudes at moderately high energies.
We then derive some of these properties
as direct consequences of the stochastic picture.
We get new results on the expression of the saturation scale of a large nucleus,
and a modified geometric scaling valid at
intermediate rapidities for dipole-dipole scattering.
\end{abstract}

\end{frontmatter}

%%%%%%%%%%%%%%%%%%%%%%%%%%%%%%%%%%%%%%%%%%%%%%%%%%%%%%%%
%%%%%%%%%%%%%%%%%%%%%%%%%%%%%%%%%%%%%%%%%%%%%%%%%%%%%%%%

\section{Introduction}

Processes such as 
the scattering of a virtual photon (which can be represented by
a distribution of color dipoles) either
off a nucleus or off another 
virtual photon
are partly described by perturbative quantum chromodynamics
when the virtuality of the photon(s) is
large enough. When the energy $\sqrt{s}$ is also high,
then the color fields generated in the
interaction are strong, and one enters an interesting 
regime in which the effect  on the scattering amplitudes
of any further
increase of the reaction energy
is described theoretically by intrinsically nonlinear
equations.
In the language of the quanta of the color field,
this is the regime in which the densities of the partons saturate.
The equation which describes saturation
is known precisely in the nucleus case in the limit
in which the number~$A$ of nucleons is very large, and in the limit
of large number~$N_c$ of colors:
It is the so-called Balitsky-Kovchegov (BK) 
equation~\cite{Balitsky:1995ub,Kovchegov:1999yj}. (The 
Jalilian~Marian-Iancu-McLerran-Weigert-Leonidov-Kovner (JIMWLK) equation 
\cite{JalilianMarian:1997jx,JalilianMarian:1997gr,Iancu:2001ad,Iancu:2000hn,Weigert:2000gi}
is a more sophisticated version of the latter, which includes finite-$N_c$ corrections).
For other processes such as $\gamma^*\gamma^*$ 
(i.e. dipole-dipole) scattering,
the relevant equations are not known for sure in the saturation
regime, but
some of their features follow from general arguments.

Some important properties of the BK equation have been understood, such
as the behavior of the saturation scale at large rapidities $y=\log (s/\Lambda_\text{QCD}^2)$, 
and the so-called ``geometric scaling'' property of the total 
deep-inelastic scattering cross section, derived
theoretically from the 
BK equation~\cite{GolecBiernat:2001if,Iancu:2002tr,Mueller:2002zm,Munier:2003vc},
after it had been discovered in the experimental data~\cite{Stasto:2000er}.

Detailed theoretical studies have been carried out in the dipole-dipole case
at ultra-asymptotic energies, and predictions for the 
rapidity dependence of the saturation scale and for the scaling
of the scattering amplitudes have been argued, based on
an analogy between what QCD evolution is expected to look like
in the saturation regime
and reaction-diffusion processes \cite{Iancu:2004es,Munier:2009pc}.
One peculiar feature of the evolution is that 
when the rapidity is large enough, no memory is kept of the
initial condition and of the early stages.

However, phenomenological analysis of the 
available experimental data have pointed out that
at realistic energies, the 
ultra-asymptotic regime may not have been reached \cite{Kozlov:2007wm}.
In this case, the initial stages of the evolution
would instead play a crucial role.

In this Letter, we shall come back to the moderate-rapidity
form of the scattering amplitude of a dipole with a nucleus (described by the
BK equation), and
investigate the case of the scattering of two dipoles.
By moderate rapidities we mean that $y$ should
be parametrically much less than $\log^3 (1/\alpha_s^2)$.
Our goal is not to build a model which can be compared right away to the data,
but to propose a picture of dipole-nucleus versus dipole-dipole scattering
at these intermediate rapidities. This picture leads to new asymptotic
formulae for the shape of the amplitude and for the $y$- and
$\alpha_s$-dependences of the saturation scale.
The limits in which our exact results are expected to be valid
($\alpha_s$ very small and fixed, $y$ very large)
are unrealistic for a direct comparison to experimental data, but we hope
that our work may pave the way to more detailed phenomenological studies.

In the next section (Sec.~\ref{sec:bk}), 
we revisit the interpretation of the
Balitsky-Kovchegov equation in different frames. We then study the
statistical properties of (linear) dipole evolution (Sec.~\ref{sec:n}), 
to finally arrive at
predictions for the parametric form of the saturation scale in
dipole-nucleus scattering and for the scaling of the amplitude in
dipole-dipole scattering (Sec.~\ref{sec:sat}).

%%%%%%%%%%%%%%%%%%%%%%%%%%%%%%%%%%%%%%%%%%%%%%%%%%%%%%%%%

\section{\label{sec:bk}Dipole-nucleus scattering viewed in different frames}

In this section,
we shall consider the scattering of a color dipole (e.g. a heavy quarkonium, or
a virtual photon) off a large nucleus.
We first recall how the BK equation is obtained from the QCD dipole model,
and how it can be replaced by the simpler FKPP equation.
Viewing the scattering process in particular frames,
we then express the dipole-nucleus scattering amplitude $T$ with the help
of $T$ itself at a lower rapidity, and of the dipole
number density $n$ obtained after dipole evolution.

\subsection{\label{sec:BK/FKPP}BK and FKPP equations}

Let us first view this process in the restframe of the nucleus, in which
the dipole is highly boosted, and thus
appears at the time of the interaction
in a high-occupancy Fock state.
The way how the Fock state of an initial dipole builds up through
the successive emissions of gluons as its
rapidity increases is conveniently described 
by the color dipole model~\cite{Mueller:1993rr}:
In the large-$N_c$ limit, gluons are
similar to $q\bar q$ pairs of zero size, and a gluon emission is interpreted 
as the splitting of a dipole into two dipoles of different sizes.
Dipole evolution is a branching-diffusion process: As the rapidity is increased by $dy$,
a dipole of size $r$ ($r$ is a 2-dimensional vector)
may be replaced by two new dipoles of respective sizes
$r^\prime$ and $r-r^\prime$ with probability
\be
\bar\alpha dy\,\frac{d^2 r^\prime}{2\pi}
\frac{r^2}{r^{\prime 2}(r-r^\prime)^2},
\label{eq:dipolekernel}
\ee
where $\bar\alpha\equiv \alpha_s N_c/\pi$.
We shall be concerned with the dipole
density {\it at a fixed impact parameter}
since this is what is relevant in scattering problems. It is very important 
to keep in mind that
under this condition, there is a largest and a smallest dipole in each realization
of the evolution, whereas if we considered all impact parameters simultaneously,
the evolution would generate an infinity of dipoles of arbitrarily small sizes.
(Later, we will replace
the full QCD dipole evolution by the simplest branching 
diffusion process in which this property will be built in).

The evolution with $y$ of the $S$-matrix element for the elastic scattering of 
an elementary dipole of size $r$ off a target such as a large nucleus can easily be
deduced from this probability distribution.
It is given by the BK equation \cite{Balitsky:1995ub,Kovchegov:1999yj}
\be
\frac{\partial}{\partial y} S(y,r)
=\bar\alpha\int \frac{d^2 r^\prime}{2\pi}
\frac{r^2}{r^{\prime 2}(r-r^\prime)^2}
\left[
S(y,r^\prime)S(y,r-r^\prime)
-S(y,r)
\right].
\label{eq:BK}
\ee
The easiest way to establish this equation is to start from the restframe
of the dipole, in which the nucleus has the rapidity $y$, and write the
change in $S$ induced when the dipole
is boosted by $dy$. The initial condition will be discussed later.

The physical picture of this mathematical description 
in the form of a {\it deterministic}
integro-differential equation is clearest in the restframe of the dipole,
in which the whole evolution takes place in the nucleus:
The nucleus being a compound of many independent nucleons
from the beginning, 
the evolution of its scattering amplitude with a probe should essentially be deterministic,
at least for small up to moderate rapidities,
for a mean-field or a classical approximation is justified by the large number
of objects.
The nonlinearity present in Eq.~(\ref{eq:BK}) is a unitarity-preserving term,
which makes sure that $0\leq S\leq 1$ throughout the evolution.
Note that there is no explicit nonlinear effect in the dipole evolution
completely determined by Eq.~(\ref{eq:dipolekernel}):
We shall add saturation in the dipole wavefunction in Sec.~\ref{sec:sat}.

For the sake of simplifying the discussion,
we first observe that due to the form of the kernel, the appropriate scale for the dipole sizes
$r$ is actually a logarithmic scale, hence in the following, we will replace
$r$ by the variable $x\equiv\log (r^2/r_0^2)$, where $r_0$ is an arbitrary size which
we shall choose later. 
We define the number density $n(y,x^\prime)$ 
of dipoles of
logarithmic size $x^\prime$ at rapidity $y$,
starting from a single dipole at $x=0$.
The manifest scale invariance of the evolution kernel~(\ref{eq:dipolekernel}) 
in the $r$ variable
becomes a translation invariance in the $x$ variable:
Therefore, the number density of dipoles  starting
with some generic $x$ is just $n(y,x^\prime-x)$.

Whenever the explicit form of the evolution
is needed, instead of attempting to deal with the full dipole evolution,
we shall replace it by the simplest possible
branching random walk (BRW): When the rapidity is increased by the infinitesimal
quantity $dy$, each given dipole characterized say 
by the variable $x$ may split to two dipoles
at $x$ with probability $dy$, and may diffuse in $x$.
The first process is the dipole branching, 
the diffusion accounts for the fact that
when a dipole splits, its offspring actually have different sizes.
In this framework, the equivalent of the BK equation~(\ref{eq:BK}) 
is the Fisher-Kolmogorov-Petrovsky-Piscounov
(FKPP) equation \cite{F:1937,KPP:1937,Munier:2003vc}, namely
\be
\partial_y S(y,x)=\partial_x^2 S(y,x)-S(y,x)+\left[S(y,x)\right]^2.
\label{eq:fkpp}
\ee
In the original FKPP equation, $y$ is the time, and $x$ a spatial variable:
Therefore, from now on, we shall often call the $x$-variable 
``position''.

The basic reason why we can afford to replace dipole branching by a simpler model
is that the solutions to the BK/FKPP equation are to a large extent universal,
namely independent of the details.
Generally speaking, at large rapidity, $S$ tends to a traveling wave, namely
a front translating as rapidity increases while keeping its shape 
essentially unchanged.
Mathematically, this means that at large $y$, $S(y,x)$ 
becomes a function of $x-\tilde X_y$ only.
(The $y$-dependence of the position of the wave front
$\tilde X_y$ will be specified later on.)
What is important to recall at this stage is that $\tilde X_y$ and the shape
of $1-S(y,x)$, whose asymptotic expression for ${x-\tilde X_y}$
large and negative reads $e^{\gamma_0(x-\tilde X_y)}$,
do not depend on the details of the initial condition, provided
that the latter is steep enough, namely that $1-S(y=0,x)\sim e^{\beta x}$ 
with $\beta>\gamma_0$.
The parameter $\gamma_0$ is determined by the linearized part of the BK/FKPP equation,
and its numerical value is $0.63\cdots$ in the case of the BK equation, and 1 for the 
FKPP equation.
The other few parameters which characterize the subasymptotic 
corrections to the shape of the front and the position
of the traveling wave are also determined by the linearized part of the
evolution equation, and may easily be replaced when changing model.

%%%%%%

\subsection{Expression for the S-matrix element in different frames}

\begin{figure}
\begin{center}
\includegraphics[width=11cm,angle=0]{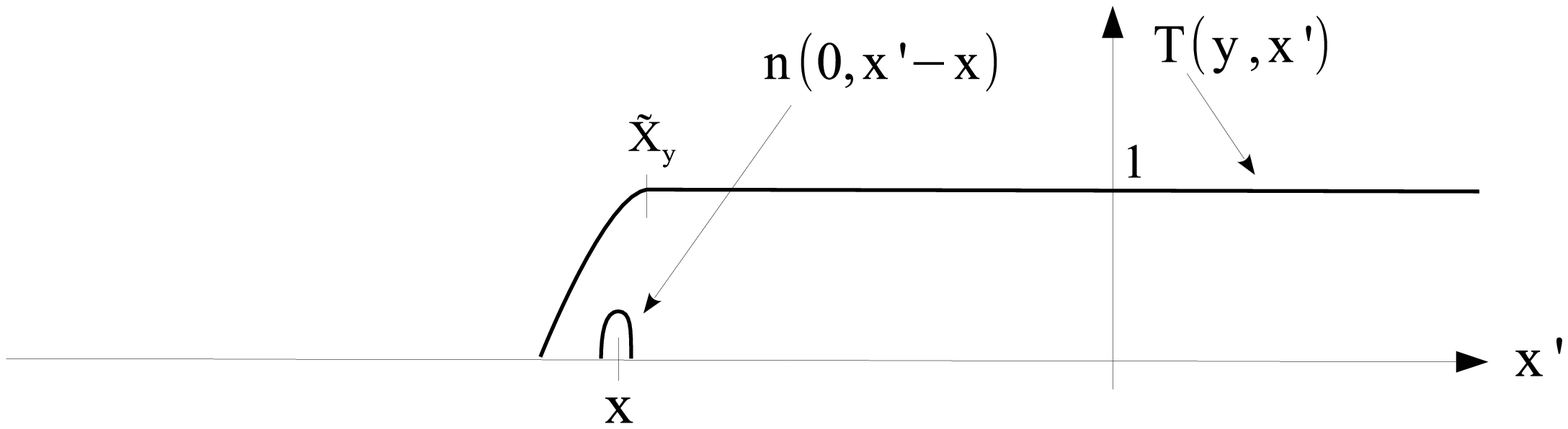}\\
\vskip 1cm
\includegraphics[width=11cm,angle=0]{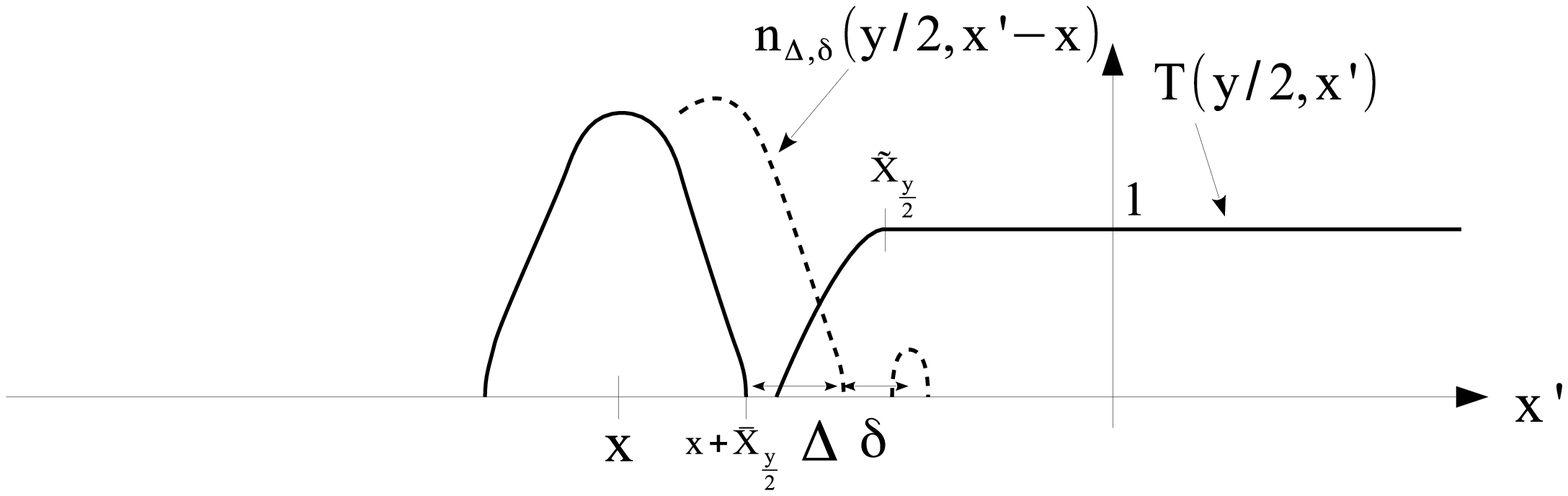}\\
\vskip 1cm
\includegraphics[width=11cm,angle=0]{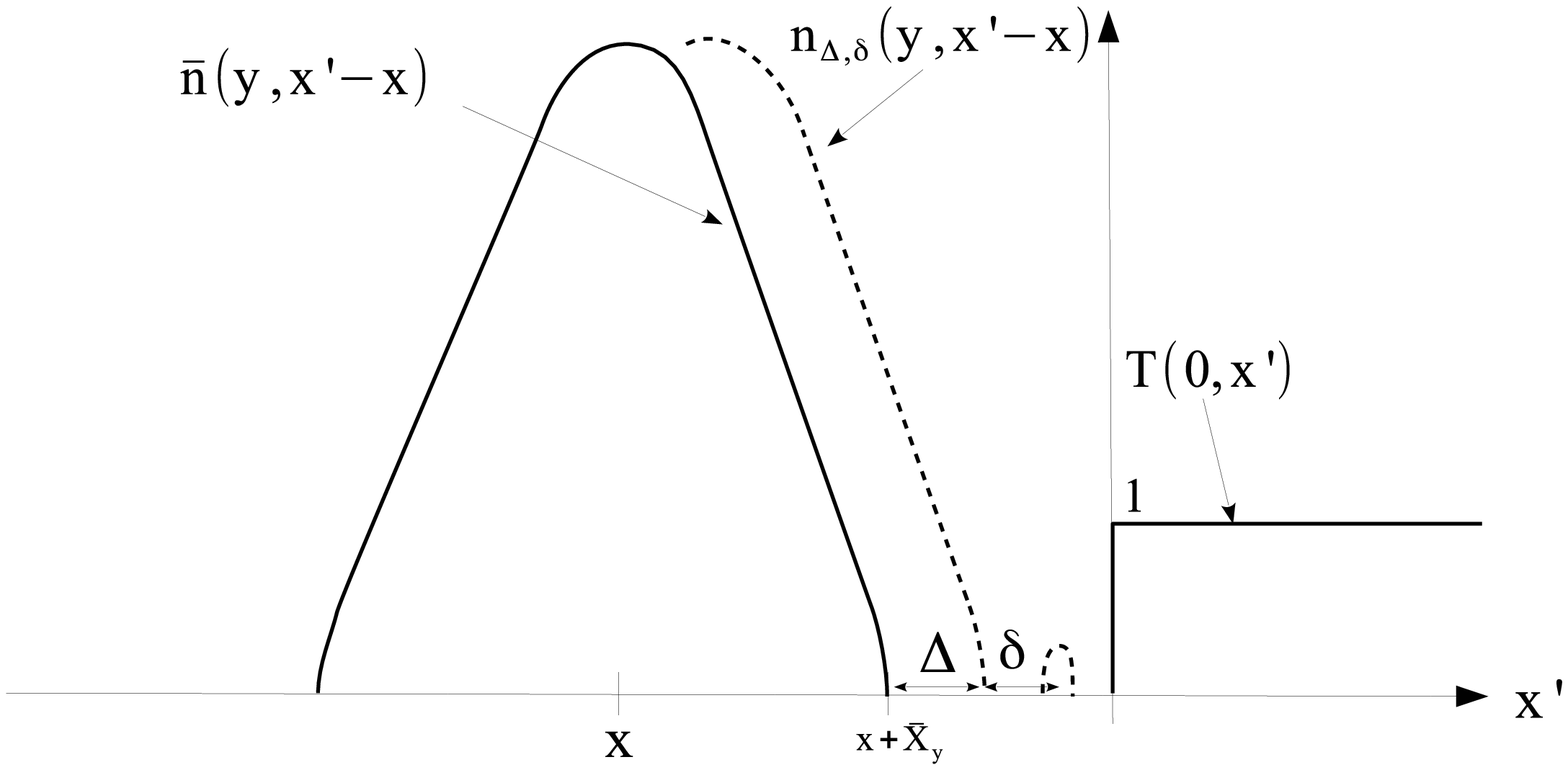}
\end{center}
\caption{\label{fig:dA}
Schematic picture of the states of the dipole and of the nucleus 
at the time of the scattering in one particular realization of the evolution, 
as viewed 
in three different frames.
The nucleus is represented by its scattering amplitude $T$ with an
elementary dipole, whose evolution obeys the BK/FKPP equation,
while the dipole evolves stochastically according to the model
described in Sec.~\ref{sec:n}.
{\it Top: Dipole restframe.} All the evolution is in the nucleus: 
The scattering amplitude is a left-moving traveling wave solution 
of the BK/FKPP equation. In this frame, the evolution is fully deterministic: This means that
the measured amplitude is merely the function $T(y,x^\prime=x)$.
{\it Middle: Center-of-mass frame.} The evolution is equally 
shared between the dipole and the nucleus. The rapidity evolution
replaces the initial elementary dipole at position $x$ 
by a stochastic density $n_{\Delta,\delta}(y/2,x^\prime-x)$. 
The measured amplitude would be the convolution of $n$ and $T$,
averaged over the realizations of the dipole evolution.
{\it Bottom: Nucleus restframe.}
The nucleus remains in its initial state.
In our idealized model for the nucleus (Eq.~(\ref{eq:SMVs})), 
the measured amplitude would simply correspond to the fraction
of realizations for which $x+\bar X_y+\Delta+\delta$ is positive, namely for
which there is an overlap between $n$ and $T$. (In the particular event represented here,
there is no such overlap).
}
\end{figure}

Let us write $S$ in a generic frame in which the rapidity is shared between
the dipole and the nucleus.
We boost the dipole to the rapidity $y_0$, keeping the total rapidity fixed at $y$.
Then at the time of the interaction, the elementary dipole
initially at position $x$
has fluctuated into a random set of dipoles of number density $n(y_0,x^\prime-x)$ at
position $x^\prime$.
We assume that these dipoles interact independently of each other with the target,
which is the key assumption leading to the BK equation.
Let us view the variable
$x^\prime$ as discretized in bins of (infinitesimal) size $dx^\prime$.
Then the $S$-matrix element for the scattering 
of a dipole at position $x$ off the nucleus
reads, at rapidity $y$,
\be
S(y,x)=\left\langle
\prod_{x^\prime}
\left[
S(y-y_0,x^\prime)
\right]^{n(y_0,x^\prime-x)dx^\prime}
\right\rangle,
\label{eq:Sgeneric}
\ee
with the convention $0^0=1$.
The average is taken over the realizations of the evolution
of the dipole, namely, in a particle physics language, over events.

If $y_0=0$, the above equality is trivial, since we are back to the 
restframe of the dipole in which $n(0,x^\prime)=\delta_{x^\prime,0}$ 
(see Fig.~\ref{fig:dA}, top).
Let us go instead to the restframe of the nucleus
by setting $y_0=y$ (see Fig.~\ref{fig:dA}, bottom).
$S(0,x^\prime)$ which appears in the r.h.s. of Eq.~(\ref{eq:Sgeneric})
represents the scattering matrix element
of a dipole at position $x^\prime$
off a large nucleus at
zero rapidity, and this is given by the 
McLerran-Venugopalan model \cite{McLerran:1993ni,McLerran:1993ka,McLerran:1994vd}.
In $r$ space, it reads
\be
S(y=0,r)=S_\text{MV}(r)=e^{-\frac{r^2 Q_\text{MV}^2}{4}\log \frac{1}{r\Lambda_{\text{QCD}}}}.
\label{eq:SMV}
\ee
$Q_\text{MV}$ is the saturation momentum of the nucleus
(It depends on the number of nucleons and on the parton density in each of them.)
$S_\text{MV}$ is rapidly going to 1 as soon as $|r|$ becomes
smaller than $2/Q_\text{MV}$, and is 0 for $|r|\gg 2/Q_\text{MV}$.
For small $r$, neglecting the subleading log factor and some uninteresting constants,
$1-S_\text{MV}(r)\sim r^2 Q_\text{MV}^2$, which is proportional
to $e^x$ in logarithmic variables. This is steeper than $e^{\gamma_0 x}$
and thus, according to the theory of traveling waves (see Sec.~\ref{sec:BK/FKPP}), 
it should not make a significant difference
to replace $S_\text{MV}$ by a step function in the $x$ variable
in the context of QCD where $\gamma_0<1$. Thus we shall opt for the
following simplified form for~$S$:
\be
S(y=0,x)=\theta\left(-x\right),
\label{eq:SMVs}
\ee
where we have set the scale $r_0$ of the transverse sizes
to twice the inverse saturation
momentum  of the nucleus: $r_0=2/Q_\text{MV}$.
The physical meaning of Eqs.~(\ref{eq:SMV}),(\ref{eq:SMVs}) 
is obvious: Dipoles which have $x>0$, 
namely sizes $|r|$ larger than
the inverse saturation scale of the nucleus are absorbed, while the nucleus
is transparent to dipoles of smaller sizes.
Inserting Eq.~(\ref{eq:SMVs}) into Eq.~(\ref{eq:Sgeneric}),
\be
S(y,x)=\left\langle
\prod_{x^\prime}
\left[
\theta\left(-x^\prime\right)
\right]^{n(y,x^\prime-x)dx^\prime}
\right\rangle.
\label{eq:Slab0}
\ee
This equation literally
means that 
\be
S(y,x)=\left(
\text{\begin{minipage}{8cm}
\begin{center}
probability that 
all dipoles sit at a position $x^\prime<0$
after evolution of a single dipole 
initially
at position $x$ 
for $y$ units of rapidity
\end{center}
\end{minipage}}
\right).
\label{eq:Slab}
\ee
Hence
\be
P(y,X)\equiv \frac{\partial S(y,-X)}{\partial X}
\label{eq:defP}
\ee
is the distribution of the position $X$
of the rightmost
particle, namely of the logarithmic
size of the largest dipole, in a BRW
which starts with a dipole at
$x=0$, and which undergoes evolution 
for $y$ units of rapidity.

We now move to the center-of-mass frame in which the rapidity
is equally shared between the dipole and the nucleus: $y_0=y/2$ 
(see Fig.~\ref{fig:dA}, middle).
We may rewrite $S$ in Eq.~(\ref{eq:Sgeneric}) in the following way:
\be
S(y,x)=\left\langle
\exp
\left[
\int dx^\prime\,
{n(y/2,x^\prime-x)
\log S(y/2,x^\prime)
}
\right]
\right\rangle.
\ee
We observe that the values of $S$ which effectively contribute
to the r.h.s. are $S\sim 1$. Therefore,
we can expand
 $\log S\equiv\log (1-T)\sim -T$ in the integrand.
We arrive at the expression
\be
S(y,x)=1-T(y,x)=
\left\langle
\exp
\left[
-\int dx^\prime\,
{n(y/2,x^\prime-x)
T(y/2,x^\prime)
}
\right]
\right\rangle.
\label{eq:Scom}
\ee
Both in the right-hand and left-hand sides of this equation,
$S=1-T$ is a solution to the FKPP equation~(\ref{eq:fkpp})
with the initial condition~(\ref{eq:SMVs}), 
namely a left-moving traveling wave.
Here again, the average $\langle\cdots\rangle$ is on the realizations
of the dipole evolution, which generates a stochastic density
of dipoles $n(y/2,x^\prime-x)$ at rapidity $y/2$ starting with a
single dipole at position $x$,
while $T$ represents the nucleus whose evolution is assumed to
be deterministic.

In the region of interest $T\ll 1$ and for large enough rapidities, 
this solution reads~\cite{BD:1997}
\be
T(y,x)=C_T
\left(
\tilde X_y-x
\right)
\exp
\left[
x-\tilde X_y-\frac{\left(x-\tilde X_y\right)^2}{4y}
\right]
\theta\left(
\tilde X_y-x
\right),
\label{eq:Tcutoff}
\ee
where
\be
\tilde X_y =-2y+\frac32\log y
\label{eq:Xtilde}
\ee
is, up to a constant of order 1, the large-$y$ expression for
the position of the FKPP front, namely the smallest $x$ for which $T$
is larger than say $\frac12$.
$C_T$ is a constant of order 1.
Equation~(\ref{eq:Tcutoff}) is valid for $y\gg 1$ and
$\tilde X_y-2\sqrt{y}< x< \tilde X_y$.
We see from Eq.~(\ref{eq:Xtilde}) (see also Fig.~\ref{fig:dA})
that the front is left-moving on the $x$-axis:
Indeed, smaller values of $x$ correspond to smaller dipoles, and the saturation
momentum must indeed move to larger momenta as rapidity increases.

We now need a model for the distribution of the dipole size density $n$.
This is the subject of the next section.

%%%%%%%%%%%%%%%%%%%%%%%%%%%%%%%%%%%%%%%%%%%%%%%%%%%%%%%%%

\section{\label{sec:n}Dipole number density and its fluctuations}

\subsection{Picture of the dipole evolution and stochastic model for $n$}

We start the dipole branching-diffusion process with a single dipole at $x=0$.

For small $y\sim 1$, the density of dipoles at position $x$ and rapidity $y$,
$n(y,x)$, is very noisy
due to the small number of objects. At large $y\gg 1$, a 
smooth distribution
builds up around $x=0$ since the typical
number of dipoles increases exponentially
with $y$, allowing for a mean-field approximation for the evolution.
The tails at large $|x|\sim 2y$ where the particle density is
low remain noisy instead, but the effect of this statistical noise
may be taken into account in a first approximation by the so-called
Brunet-Derrida cutoff \cite{BD:1997}, which is a moving absorptive boundary.
It is actually the rightmost tail of the distribution of dipoles
which is relevant to the computation of the scattering amplitude,
see Fig.~\ref{fig:dA}.
The solution of the deterministic evolution of the dipoles
with this cutoff enforcing discreteness reads, near the 
rightmost boundary (located at position $x=\bar X_y\sim +2y$),
\be
\bar n(y,x)=C_{\bar n}(\bar X_y-x)
\exp
\left[
\bar X_y-x-\frac{\left(x-\bar X_y\right)^2}{4y}
\right]
\theta\left(
\bar X_y-x
\right),
\label{eq:ncutoff}
\ee
where up to a constant of order one, the position of the boundary is
\be
\bar X_y=2y-\frac32 \log y.
\label{eq:Xbar}
\ee
Equation~(\ref{eq:ncutoff}) is valid for $\bar X_y-2\sqrt{y}<x<\bar X_y$.

Note that the $y$-dependence of $\bar X_y$ is precisely the same as for the
position of the FKPP traveling wave, see the expression of $-\tilde X_y$ 
in Eq.~(\ref{eq:Xtilde}).
Technically, this is clear since in order to get these expressions,
in both cases, one puts an absorptive boundary on
a linear branching-diffusion equation, see e.g. Ref.~\cite{Munier:2009pc}.
More deeply, this identity between $\bar X_y$ and $-\tilde X_y$ actually is
a duality of the FKPP equation, see the mathematical work of Ref.~\cite{doering:2003}
and the recent related work of ours~\cite{Mueller:2014gpa}.

We shall now propose a model for
the fluctuations that deform this solution.
They may occur in two different places. First, as already mentioned,
in the early stages of the evolution, the whole
system is stochastic since the overall number of dipoles is small. 
After further rapidity evolution,
the early fluctuations essentially result
in fluctuations of the position of the boundary $\bar X$ 
of the deterministic form~(\ref{eq:ncutoff})
by some random $\Delta$,
where $\Delta$ has an a priori rapidity-dependent distribution, 
which we shall denote by $p_f(y,\Delta)$.
We call these fluctuations ``front fluctuations''.
At rapidities $y\gg 1$, 
when the total number of dipoles
is large, fluctuations still occur
near the tip of the distribution.
These tip fluctuations consist in sending randomly a small number of particles
ahead of the deterministic front by some distance $\delta$, which has 
the distribution $p_t(\delta)$ to be determined later. 
The simplest model for the shape of 
these fluctuations is a Dirac
distribution $\delta_D$ with support
at position $\bar X_y+\Delta+\delta$.
We call these fluctuations ``tip fluctuations''.

We write
\begin{multline}
n_{\Delta,\delta}(y,x)=\bar n(y,x-\Delta)+C\times \delta_D(x-\bar X_y-\Delta-\delta)\\
\text{ with probability $\left[p_f(y,\Delta)d\Delta\right]
\left[p_t(\delta)d\delta\right]$},
\label{eq:nmodel}
\end{multline}
where $C$ is a constant of order 1 which encodes our very ignorance
of the detailed shape of the forward fluctuations.
A sketch of the evolution
of $n$ in this model is represented in Fig.~\ref{fig:dA} (middle and bottom).

We refer the reader to the recent paper of Ref.~\cite{Mueller:2014gpa}
for a more complete discussion of the fluctuations in a general branching
random walk.

\subsection{Constraining the distributions of fluctuations}

Interestingly enough, 
we can actually to a large extent ``guess'' 
the distributions $p_f$ and $p_t$
of the two kinds of fluctuations we have identified.
To this aim, 
we take a generating function of the moments of $P$ defined in Eq.~(\ref{eq:defP}), 
namely of the moments of the
distribution of the position $X$ of the rightmost particle in the BRW:
\be
\left\langle
e^{\lambda X}
\right\rangle_y
=\int_{-\infty}^{+\infty}dX e^{\lambda X}
P(y,X)
=\int_{-\infty}^{+\infty}dX e^{\lambda X}
\frac{\partial S(y,-X)}{\partial X}.
\label{eq:genmom}
\ee
The $y$-index for the expectation value is meant to keep track of the fact
that $X$ has a $y$-dependent probability distribution.

We then go to the restframe of the nucleus
in which $S$ is related to $n$ through Eq.~(\ref{eq:Slab0}).
Using the model~(\ref{eq:nmodel}) for $n$, we get
\be
\begin{split}
S(y,x)&=
\left\langle
\theta(-x-\bar X_y-\Delta-\delta)
\right\rangle_y\\
&=
\int d\Delta\, p_f(y,\Delta)\int d\delta\, p_t(\delta)\,\theta(-x-\bar X_y-\Delta-\delta).
\end{split}
\label{eq:Slabmodel}
\ee
Inserting Eq.~(\ref{eq:Slabmodel}) into Eq.~(\ref{eq:genmom}), 
a straightforward
calculation leads to the following relation between generating functions
of centered moments:
\be
\left\langle
e^{\lambda (X-\langle X\rangle_y)}
\right\rangle_y
=
\left\langle
e^{\lambda (\Delta-\langle\Delta\rangle_y)}
\right\rangle_y
\left\langle
e^{\lambda (\delta-\langle \delta\rangle)}
\right\rangle.
\label{eq:FmomA}
\ee
As always, the averages are over realizations of the dipole evolution,
and the index $y$ keeps track of the rapidity at which the mean is taken.

Of course, the factorization in the r.h.s. 
of Eq.~(\ref{eq:FmomA})
just follows from the assumption that
the front fluctuations $\Delta$ and the tip fluctuations
$\delta$ are uncorrelated, which should
be true for large enough values of the rapidity.

We now move to the center-of-mass frame, in which $S$ is given by Eq.~(\ref{eq:Scom}).
We insert Eq.~(\ref{eq:Tcutoff}) and the model~(\ref{eq:nmodel}) into~(\ref{eq:Scom}),
and perform the integral over $x^\prime$. Keeping the leading term when $y\gg 1$, we find
for this integral
\be
\int dx^\prime {n_{\Delta,\delta}(y/2,x^\prime-x)
T(y/2,x^\prime)
}\simeq C_{\bar n}C_T\frac{\sqrt{\pi}}{4}
y^{3/2} e^{\bar X_{y/2}-\tilde X_{y/2}+x+\Delta}.
\label{eq:int}
\ee
There is no $\delta$ dependence in the r.h.s., since the tip fluctuations
would bring a negligible contribution
to the integral over $x^\prime$. 
Using the expressions~(\ref{eq:Xtilde}) and~(\ref{eq:Xbar}) for $\tilde X$ 
and $\bar X$ respectively,
$S$ may be written as
\be
S(y,x)=\left\langle
\exp\left(
-\alpha e^{\bar X_{y}+x+\Delta}
\right)
\right\rangle_y,
\label{eq:Scom2}
\ee
with $\alpha$ a constant of order 1 which includes the constants in Eq.~(\ref{eq:int})
and the unknown additive constants in $\bar X_y$.
We take again the moments of $P$ starting from Eq.~(\ref{eq:Scom2}).
We find
\be
\left\langle
e^{\lambda (X-\langle X\rangle_y)}
\right\rangle_y
=\Gamma(1-\lambda)e^{-\gamma_E\lambda}
\left\langle
e^{\lambda (\Delta-\langle\Delta\rangle_y)}
\right\rangle_y.
\label{eq:Fmomcom}
\ee
Identifying Eq.~(\ref{eq:Fmomcom}) to Eq.~(\ref{eq:FmomA}),
we can get the generating function of the tip fluctuations:
\be
\left\langle
e^{\lambda (\delta-\langle\delta\rangle)}
\right\rangle=\Gamma(1-\lambda)e^{-\gamma_E\lambda}.
\ee
Hence the probability distribution of $\delta$
is a Gumbel distribution:
\be
p_t(\delta)= c\exp\left(-\delta-c e^{-\delta}\right),
\ee
where $c$ is a constant of order 1,
which may easily
be expressed with the help of $\langle\delta\rangle$.

Incidentally, it seems that we have recovered the Lalley and Sellke
theorem \cite{LS:1987} for the fluctuations of the boundary
of a branching random walk, provided that $\Delta$
be identified to the random variable $\log Z_y$, with
\be
Z_y=
\sum_i \left[2y-x_i(y)\right]
e^{x_i(y)-2y},
\ee
where $x_i(y)$ is the position of particle $i$ in a particular realization
of the evolution at rapidity $y$, and the sum goes over the particles present
at this same rapidity.

Finally, in the restframe of the dipole, $S(y,-X)$ is simply related
to the solution of the FKPP equation.
We do not have a full analytic form for this solution, however
$S(y,-X)$ can be deduced from Eq.~(\ref{eq:Tcutoff}) for $X+\tilde X_y\gg 1$
and large $y$.
This turns out to be enough to enable us to evaluate the 
generating function of the moments of $X$
in the limit $\lambda\rightarrow 1$ in which
the integral over $X$ in Eq.~(\ref{eq:genmom}) is dominated by large values of $X$.
Integrating Eq.~(\ref{eq:genmom}) by parts for $0<\lambda<1$,
\be
\left\langle e^{\lambda X}\right\rangle_y=
\lambda\int_{-\infty}^{+\infty}dX e^{\lambda X}T(y,-X).
\ee
At large rapidities and keeping the leading singularity when $\lambda\rightarrow 1$,
\be
\left\langle e^{\lambda (X+\tilde X_y)}\right\rangle_y
\underset{\lambda\to 1,
y\to +\infty}{\longrightarrow}
\frac{C_T}{(1-\lambda)^2}.
\ee
Noticing that 
$\tilde X_y$ is, up to a sign and to a constant, equal to the
average position of the rightmost particle in the BRW, namely
$\tilde X_y=-\langle X\rangle_y+\text{const}$, we
may identify this expression to Eq.~(\ref{eq:Fmomcom}). 
We see that the generating function 
of the moments of $\Delta$ must have a simple
pole at $\lambda=1$, which means that
\be
p_f(y,\Delta)\underset{1\ll\Delta\ll\sqrt{y}}{\sim} e^{-\Delta}.
\label{eq:pf}
\ee
We expect finite-$y$ corrections: The exponential tail
must be cut off at a distance $\Delta\sim\sqrt{y}$, but
this limitation is irrelevant at large $y$ since obviously,
typical $\Delta$ are
of order 1.

We further note that the identification of the $S$-matrix element
in the dipole restframe
with the same quantity in the nucleus restframe
enables one to relate
the shape of the traveling wave solution of the FKPP equation
to the fluctuations occurring in the initial stages of the rapidity
evolution of the dipoles, as seen from the equation
\be
\int_{-\infty}^{+\infty}
dX\, e^{\lambda\left(X-\langle X\rangle_y\right)}
\left[
-\frac{\partial T(y,-X)}{\partial X}
\right]
=\Gamma(1-\lambda)e^{-\gamma_E\lambda}
\left\langle
e^{\lambda(\Delta-\langle \Delta\rangle_y)}
\right\rangle_y.
\ee
The tip fluctuations are represented by the factor $\Gamma(1-\lambda)$
in the r.h.s.
($e^{-\gamma_E\lambda}$ is a mere normalization factor).
This equation says that the shape of a BK/FKPP traveling wave near the
unitarity region is directly
related to the front fluctuations, that is, to the initial stages
of the evolution.

%%%%%%%%%%%%%%%%%%%%%%%%%%%%%%%%%%%%%%%%%%%%%%%%%%%%%%%%%

\section{\label{sec:sat}
Including saturation in the dipole evolution:
Predictions for amplitudes at moderate rapidities
}

\begin{figure}
\begin{center}
\includegraphics[width=11cm,angle=0]{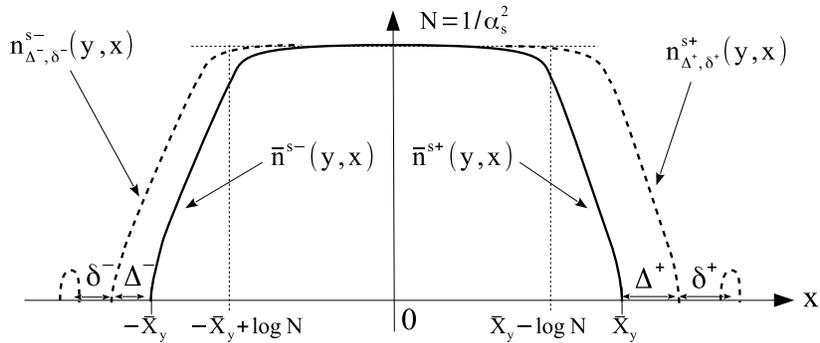}
\end{center}
\caption{\label{fig:dsat}
Schematic picture of a realization
of the density of dipoles at rapidity $y$
starting with a dipole at position $x=0$.
$\bar X_y$ is the position of the right discreteness cutoff,
$\bar X_y-\log N$ is the point where nonlinear saturation effects
start to be significant, and thus the location of the right saturation boundary.
The analytical expressions~(\ref{eq:barns+}),(\ref{eq:barns-}) 
are valid within these boundaries.
}
\end{figure}

So far, we have treated the evolution of the dipole as a branching process
(with diffusion in the transverse momentum) with rate independent
of the dipole density.
No nonlinear mechanism was included in the evolution.
(The nonlinearity in the BK equation may be seen as due to the
independent multiple scatterings
of the set of dipoles present in the wavefunction at rapidity $y$.)

There are however convincing arguments to expect
that at higher energies, the growth of the gluon/dipole number density
must slow down.
This should happen in the phase space regions
where the number density of dipoles becomes as large 
as $N\equiv 1/\alpha_s^2$.
At a rapidity $y\gg\log 1/\alpha_s^2$, the dipole density should look like
the sketch in Fig.~\ref{fig:dsat}. We shall call it $n^s$, and discuss its analytical
properties before we use it to compute the dipole-nucleus and dipole-dipole
scattering amplitudes.

\subsection{Dipole number density with saturation}

In practice, saturation can be implemented in the form of moving absorptive boundaries
making sure that $n\leq N$ at all rapidities \cite{Mueller:2004sea}.
These ``unitarity'' boundaries turn out to be located
at a distance $\pm\log N$ of the discreteness
boundaries.

The effect of saturation in the dipole evolution is to modify the
shape of the dipole density, and the $y$-dependence of the position
of the discreteness cutoff \cite{BD:1997}.
Starting with a single dipole at position 0, 
the position of the right discreteness cutoff now reads
\be
\bar X_y=\begin{cases}
2y-\frac32\log y & \text{for $y\ll\log^2 N$},\\
\left(2-\frac{\pi^2}{\log^2 N}\right)y-3\log \log N & \text{for $y\gg\log^2 N$},
\end{cases}
\label{eq:barXsat}
\ee
up to constants of order one.
The left cutoff is at position $-\bar X_y$.
As for the shape of the particle density,
in a first approximation, for $y\gg \log^2 N$,
$\bar n$ in Eq.~(\ref{eq:ncutoff})
is replaced by \cite{BD:1997}
\begin{multline}
\bar n^{s+}(y,x)=
C_{\bar n^s}\frac{\log N}{\pi}\left[\sin\frac{\pi(\bar X_y-x)}{\log N}\right]
\exp\left(\bar X_y-x\right)\\
\times\theta(\bar X_y-\log N<x<\bar X_y).
\label{eq:barns+}
\end{multline}
The $\theta$ function indicates that this formula is valid
within a distance $\log N$ of the right discreteness boundary.

We also write the expression of the particle density near the left discreteness
boundary:
\begin{multline}
\bar n^{s-}(y,x)=
C_{\bar n^s}\frac{\log N}{\pi}\left[\sin\frac{\pi(\bar X_y+x)}{\log N}\right]
\exp\left(\bar X_y+x\right)\\
\times\theta(-\bar X_y<x<-\bar X_y+\log N).
\label{eq:barns-}
\end{multline}
These smooth functions can be promoted to stochastic functions $n^{s\pm}_{\Delta,\delta}$
by adding the front and tip fluctuations discussed before,
as in Eq.~(\ref{eq:nmodel}). It is enough to substitute $\bar n$ by $\bar n^{s\pm}$ therein:
\be
\begin{split}
n^{s+}_{\Delta^+,\delta^+}(y,x)=\bar n^{s+}(y,x&-\Delta^+)+C\times \delta_D(x-\bar X_y-\Delta^+-\delta^+)\\
&\text{with probability $\left[p_f(y,\Delta^+)d\Delta^+\right]
\left[p_t(\delta^+)d\delta^+\right]$},\\
n^{s-}_{\Delta^-,\delta^-}(y,x)=\bar n^{s-}(y,x&+\Delta^-)+C\times \delta_D(x+\bar X_y+\Delta^-+\delta^-)\\
&\text{with probability $\left[p_f(y,\Delta^-)d\Delta^-\right]
\left[p_t(\delta^-)d\delta^-\right]$}.
\end{split}
\label{eq:nsmodel}
\ee
A schematic picture of these functions is represented in Fig.~\ref{fig:dsat}.

We shall use this model to compute the scattering amplitudes $T_{dA}$
of a dipole with a nucleus, and $T_{dd}$ of two dipoles,
including saturation in the wavefunction of the dipole(s).

\subsection{Dipole-nucleus scattering}

The saturation momentum of a large nucleus is easily deduced from Eq.~(\ref{eq:barXsat}).
The simplest is to go to the nucleus restframe, and to recognize that
up to an additive numerical constant of order one,
the average logarithm of the
squared saturation scale is given by
$\bar X$. 

We have so far worked with the FKPP equation.
It is quite straightforward to generalize the universal results obtained 
for that equation to a generic 
branching-diffusion process, see e.g. Ref.~\cite{Munier:2009pc}.
We denote by $\chi(\gamma)$ the eigenvalue of the dipole kernel corresponding to
the eigenfunction $|r|^{2\gamma}$, 
and by $\gamma_0$ the solution of the equation 
$\chi(\gamma_0)=\gamma_0\chi^\prime(\gamma_0)$
(which numerically gives $\gamma_0=0.63\cdots$)
\cite{Gribov:1984tu,Mueller:2002zm}.
With this kernel and switching to the variables relevant to QCD, 
we find the following expression
for the saturation scale of the nucleus:
\be
\log \frac{Q_{s,A}^2(y)}{Q_\text{MV}^2}
=
\begin{cases}
v_0\bar\alpha y-\frac{3}{2\gamma_0} \log(\bar\alpha y)  
& \text{for $\bar\alpha y\ll\frac{1}{2\gamma_0^2\chi^{\prime\prime}(\gamma_0)}
\log^2\frac{1}{\alpha_s^2}$}\\
v_\text{BD}\bar\alpha y-\frac{3}{\gamma_0}\log\log\frac{1}{\alpha_s^2}
& \text{for $\bar\alpha y\gg\frac{1}{2\gamma_0^2\chi^{\prime\prime}(\gamma_0)}
\log^2\frac{1}{\alpha_s^2}$}\\
\end{cases}
\label{eq:QsA}
\ee
where $v_0$ is the asymptotic velocity of the BK traveling wave, and
$v_\text{BD}$ includes the effect of the Brunet-Derrida 
discreteness cutoff~\cite{BD:1997,Mueller:2004sea}:
\be
v_0=\chi^\prime(\gamma_0),\ 
v_\text{BD}=v_0-\frac{\pi^2\gamma_0\chi^{\prime\prime}(\gamma_0)}{2\log^2\frac{1}{\alpha_s^2}}.
\label{eq:v0vBD}
\ee
Equation~(\ref{eq:QsA})
corrects Eq.~(26) in Ref.~\cite{Mueller:2003bz}.

The shape of the front is of course given by 
Eq.~(\ref{eq:Tcutoff}), which exhibits
the well-known form of geometric scaling \cite{Mueller:2002zm}:
\be
T_{dA}(y,r)\sim \log\frac{1}{r^2Q_{s,A}^2(y)} \left[r^2Q_{s,A}^2(y)\right]^{\gamma_0},
\label{eq:GS}
\ee
provided that $|rQ_{s,A}(y)|\ll 1$ and
$\log^2(r^2Q_{s,A}^2(y))\ll 2\chi^{\prime\prime}(\gamma_0)\bar\alpha y$.

One could wonder what happens if one chooses to 
view the scattering in another frame, e.g.
in the restframe of the dipole.
Then, at rapidities parametrically
larger than $\log^2(1/\alpha_s^2)$,
the classical
approximation breaks down, and the
FKPP evolution must be replaced by a stochastic evolution.
The main effect of stochasticity in that equation can be represented
by an appropriate Brunet-Derrida cutoff \cite{BD:1997}.

%%%%%

\subsection{Dipole-dipole scattering in the saturation regime}

\begin{figure}
\begin{center}
\includegraphics[width=11cm,angle=0]{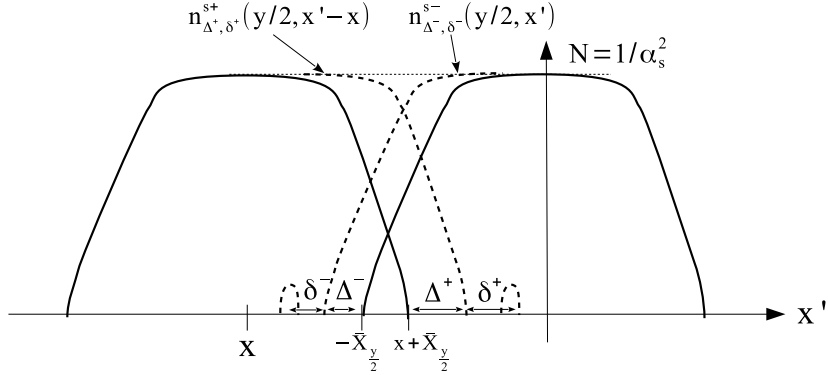}
\end{center}
\caption{\label{fig:dd}
Schematic picture of one dipole-dipole scattering event in the
center-of-mass frame at rapidity $y$, with saturation included in the
dipole evolution. What is actually represented is the density of dipoles
in the two colliding objects
after evolution over $y/2$ steps of rapidity for each object.
The scattering amplitude is given
in Eqs.~(\ref{eq:Sdd}),(\ref{eq:Tdd}) and involves in particular an average
over the realizations.
}
\end{figure}

We now consider the dipole-dipole case. 
We recall that generically, the scattering amplitude of a dipole of size
$r$ off a dipole of size $r^\prime$ at zero rapidity
is approximately local in impact parameter
and essentially reads, for two dipoles at the same impact parameter,
\be
T_{dd}^\text{el}(r,r^\prime)\sim \alpha_s^2\frac{r_<^2}{r_>^2}
\ee
where $r_<=\min(|r|,|r^\prime|)$
and $r_>=\max(|r|,|r^\prime|)$.

Here we shall take as our initial configuration
a dipole at position $x$ and another one at position 0.
Obviously, the scattering amplitude of these elementary dipoles reads
\be
T_{dd}^\text{el}(x)\sim\alpha_s^2 e^{-|x|}
\ee
and since this is an exponential steeper than $e^{-\gamma_0 |x|}$,
as in the case of the McLerran-Venugopalan model discussed above, its
width is irrelevant to the subsequent evolution
and thus the initial condition for $T_{dd}$ may be approximated by
\be
T_{dd}(y=0,x)=T_{dd}^\text{el}(x)\sim
\alpha_s^2\delta_{x,0}=\frac{1}{N}\delta_D(x).
\label{eq:Tddmodel}
\ee
After rapidity evolution, assuming that the dipoles scatter 
independently of each other,
by analogy with Eq.~(\ref{eq:Scom}),
we may write the amplitude in a general frame as (see Fig.~\ref{fig:dd}
for a sketch in the center-of-mass frame $y_0=y/2$)
\be
\begin{split}
S_{dd}(y,x)&=1-T_{dd}(y,x)\\
&=\left\langle
\exp
\left[
-\int dx^\prime\,
dx^{\prime\prime}
\,
{n^s(y_0,x^\prime-x)
T_{dd}^\text{el}(x^\prime-x^{\prime\prime})
n^s(y-y_0,x^{\prime\prime})
}
\right]
\right\rangle.
\end{split}
\label{eq:Sdd}
\ee
Assuming without loss of generality that $x<0$,
looking again at Fig.~\ref{fig:dd}, we replace the saturated dipole densities $n^s$ 
in the exponential by their appropriate form from the model in Eq.~(\ref{eq:nsmodel}), 
$T_{dd}^{\text{el}}$ by its expression in Eq.~(\ref{eq:Tddmodel}),
and we express explicitly the
average over realizations in terms of the probability distribution of the fluctuations 
given in Eq.~(\ref{eq:nsmodel}). All in all, we get
\begin{multline}
T_{dd}(y,x)=\int d\Delta^+\,p_f(\Delta^+)\int d\delta^+ \,p_t(\delta^+)
\int d\Delta^-\,p_f(\Delta^-)\int d\delta^-\, p_t(\delta^-)
\\
\times\left\{
1-\exp\left[
-\frac{1}{N}\int dx^\prime\,n^{s+}_{\Delta^+,\delta^+}(y_0,x^\prime-x)
n^{s-}_{\Delta^-,\delta^-}(y-y_0,x^\prime)
\right]
\right\}.
\label{eq:Tdd}
\end{multline}
In the restframe of the dipole sitting at 0, for $T$ much smaller than 1 
but significantly larger than $\alpha_s^2=1/N$, the above formula simplifies.
The scattering amplitude
is just the shape of the particle number density multiplied by 
the elementary dipole-dipole
amplitude, averaged over the fluctuations of the evolved dipole:
\be
T_{dd}(y,x)=
\int\, d\Delta\, p_f(y,\Delta)\int d\delta\,
p_t(\delta) 
\left[
\frac{1}{N} n^{s+}_{\Delta,\delta}(y,-x)
\right].
\ee
The tip fluctuations are irrelevant since we are looking for the scaling form
of $T$ in the region $T\gg 1/N$. They can be neglected there.

From the exponential form~(\ref{eq:pf}) 
of the probability distribution of the 
front fluctuations $\Delta$ and using Eq.~(\ref{eq:nsmodel}), we get
\be
T_{dd}(y,x)\sim (x+\bar X_y-\log N)^2 e^{x+\bar X_y-\log N}.
\label{eq:Tnewgs}
\ee
This is a new form of geometric scaling, valid in the saturation
regime at moderate rapidities, namely for $\log^2 N\ll y\ll \log^3 N$,
and this scaling is valid for $x$ satisfying $|x+\bar X_y-\log N|\ll \log N$.

It is instructive to also go to the center-of-mass frame (see Fig.~\ref{fig:dd}).
We go back to Eq.~(\ref{eq:Tdd}), set $y_0=y/2$ and expand the exponential.
Again, the tip fluctuations are negligible, but the front fluctuations of
both evolved dipoles are important:
\begin{multline}
T_{dd}(y,x)=\frac{1}{N}
\int d\Delta^+ d\Delta^-\,p_f(\Delta^+)p_f(\Delta^-)\\
\times\int dx^\prime\,\bar n^{s+}(y/2,x^\prime-x-\Delta^+)
\bar n^{s-}(y/2,x^\prime+\Delta^-).
\end{multline}
Substituting $\bar n^{s\pm}$ by Eq.~(\ref{eq:barns+}),(\ref{eq:barns-}),
performing the integration and expanding the result for $|x+\bar X_y-\log N|\ll \log N$,
we recover Eq.~(\ref{eq:Tnewgs}).

Finally, we take over Eq.~(\ref{eq:Tnewgs}) to QCD, by
substituting $\bar X_y$ by Eq.~(\ref{eq:barXsat}) and switching to the
variables relevant for QCD:
\be
T_{dd}(y,r)\sim \log^2\frac{1}{r^2Q_s^2(y)}
\left[r^2Q_s^2(y)\right]^{\gamma_0},
\label{eq:GSnew}
\ee
where the dipole saturation scale reads, for $\bar\alpha y\gg\frac{1}{2\gamma_0^2\chi^{\prime\prime}(\gamma_0)}
\log^2\frac{1}{\alpha_s^2}$,
\be
\log \left(r_0^2Q_{s}^2(y)\right)
=
v_\text{BD}\bar\alpha y-\frac{1}{\gamma_0}\log\frac{1}{\alpha_s^2}
-\frac{3}{\gamma_0}\log\log \frac{1}{\alpha_s^2}.
\label{eq:Qs}
\ee
$v_\text{BD}$ was defined in Eq.~(\ref{eq:v0vBD}).

The difference between~(\ref{eq:GSnew}) and the usual geometric scaling~(\ref{eq:GS})
is with the log which enters with a power~2 in the former. This is directly related
to the front fluctuations which build up in the early stages of the
dipole evolution.

%%%%%%%%%%%%%%%%%%%%%%%%%%%%%%%%%%%%%%%%%%%%%%%%%%%%%%%%%

\section{Summary and outlook}

In this paper, we have emphasized the role of the parton number fluctuations
especially in the initial stages of the rapidity evolution.
The importance of rare fluctuations was argued in Ref.~\cite{Iancu:2003zr}
in the context of the BK equation, but we have now a more complete
and more quantitative 
understanding of the very nature of these fluctuations.

We have derived from the stochastic picture new properties
for the scattering amplitudes when the total rapidity
is parametrically less than $\log^3 (1/\alpha_s^2)$, 
in the two following cases:

\begin{enumerate}[(i)]

\item{\it Dipole-nucleus scattering:}
The amplitude exhibits the usual geometric scaling form~(\ref{eq:GS}),
the saturation scale being given in Eq.~(\ref{eq:QsA}),

\item{\it Dipole-dipole scattering:}
The amplitude exhibits a modified geometric scaling form, given by
Eq.~(\ref{eq:GSnew}), with the saturation scale~(\ref{eq:Qs}).
This is the main new result of this paper.

\end{enumerate}

To complete the picture, let us
recall that the regime of rapidities larger than $\log^3(1/\alpha_s^2)$ was
studied before \cite{Iancu:2004es,Brunet:2005bz,Hatta:2006hs}:
The imprint of the initial stages of the evolution on the amplitude
is washed out by fluctuations occurring at a rate
$1/\log^3(1/\alpha_s^2)$, and consequently,
geometric scaling is replaced by so-called ``diffusive scaling''.

For the future, it would be interesting to test numerically 
especially the new form of geometric scaling we have found.

%%%%%%%%%%%%%%%%%%%%%%%%%%%%%%%%%%%%%%%%%%%%%%%%%%%%%%%%%
%%%%%%%%%%%%%%%%%%%%%%%%%%%%%%%%%%%%%%%%%%%%%%%%%%%%%%%%%

\section*{Acknowledgements}

We acknowledge support from ``P2IO Excellence Laboratory'',
and from the US Department of Energy.

%%%%%%%%%%%%%%%%%%%%%%%%%%%%%%%%%%%%%%%%%%%%%%%%%%%%%%%%%%%%%%%%%%%%%%%%%%%%%%%%%%%%%%

\end{document}